\documentstyle[12pt]{article}
\newcommand{\be}{\begin{equation}}
\newcommand{\bef}{\begin{figure}}
\newcommand{\eef}{\end{figure}}

\newcommand{\ee}{\end{equation}}

\begin{document} 
\title{
\begin{center}
Conceptual Problems 
\end{center}
\begin{center}
of the Standard Cosmological Model
\end{center}
}

\author{Yurij V. Baryshev \\ \\
\small Astronomical  Institute of the Saint-Petersburg University, \\
\small 198504, St.-Petersburg, Russia.  E-mail:  yuba@astro.spbu.ru}

\date{}

\maketitle


\begin{abstract}

The physics of the expansion of the universe is still a poorly studied
subject of the standard cosmological model. This because the
concept of expanding space can not be tested in the laboratory and
because ``expansion'' means continuous creation of space,
something that leads to several paradoxes.  We re-consider and expand
here the discussion of conceptual problems, already noted in the
literature, linked to the expansion of space. In particular we discuss
the problem of the violation of energy conservation for local comoving
volumes, the exact Newtonian form of the Friedmann equations, the
receding velocity of galaxies being greater than the speed of light,
and the Hubble law inside inhomogeneous galaxy distribution.  Recent
discussion by Kiang, Davis \& Lineweaver, and Whiting of the
non-Doppler nature of the Lemaitre cosmological redshift in the
standard model is just a particular consequence of the paradoxes
mentioned above.  The common cause of these paradoxes is the
geometrical description of gravity (general relativity), where there
is not a well defined concept of the energy-momentum tensor for the
gravitational field and hence no energy-momentum conservation for
matter plus gravity.

\end{abstract}

\section{The "Absurd Universe" of Modern Cosmology }


In a number of recent papers [28,29,30], one of which entitled "Absurd
universe", Michael Turner emphasized the unpleasant status of the most
widely accepted cosmological model, where about 95 \% of the total
matter density of the universe have unknown physics.  This so called
``consensus universe'' being dominated by cold dark matter and dark
energy, two hypothetical entities not at all tested in the laboratory.

Beside this problem with the ``dark sector'' of the universe, the
situation is made even worse when one considers the severe conceptual
problems of the standard cosmological model arising even within the
boundaries of the known physics. These problems were first discovered
and analyzed by Edward Harrison (1981, 1993, 1995)[13, 14, 15] but are
poorly known and very rarely discussed in the literature.  Here we
consider a number of puzzling properties of Friedmann expanding models:
the violation of energy-momentum conservation for any local comoving
ball with non-zero pressure, the exactly Newtonian form of the
relativistic Friedmann equations, the unlimited receding velocities of
galaxies, and the linear Hubble law inside strongly inhomogeneous
galaxy distribution. Existence of these paradoxes means that the
standard model of the universe is much more "absurd" than one usually
thinks.

In this report an account of the origin of these conceptual problems
is given and a common cause for their occurrence is proposed.

\section{Two-fluid matter-dark energy FLRW model}

The two basic ingredients of modern cosmological models  are:
\begin{itemize}
\item {\em relativistic  theories of gravity, and}
\item {\em the cosmological principle.}
\end{itemize}
For the Friedmann-Lemaitre-Robertson-Walker (FLRW) model, which 
is the currently accepted basis for the interpretations of 
all astrophysical observations and the basis of the
Standard Cosmological Model (SCM),
these ingredients are the general relativity and Einstein's
cosmological principle.
Modern versions of SCM make important distinction between usual matter
(with positive pressure) and dark energy (with negative pressure).
A general classification and the main properties of the two-fluid FLRW
models were recently discussed by Gromov et al.(2004)[11].

\subsection{General Relativity}

The first fundamental element of the SCM  is the
General Relativity (GR), which is a geometrical gravity theory.
GR was successfully tested in the weak gravity condition
of the Solar System and binary neutron stars.
It is assumed that GR can be applied to the Universe as a whole.

According to GR gravity is described by a metric tensor
$\:g^{ik}$ of a Riemannian space. The "field" equations in GR
(Einstein-Hilbert equations) have the form
(we use Landau \& Lifshitz 1971 [19] notations):
\begin{eqnarray}
\Re^{ik} - \frac{1}{2}\,g^{ik}\,\Re =
\frac{8\,\pi\,G}{c^4}\,\,
\left(T^{ik}_{(m)} + T^{ik}_{(de)}\right)
\label{eheq}
\end{eqnarray}
where $\:\Re^{ik}$ is the Ricci tensor,
$\:T^{ik}_{(m)}$ is the energy-momentum tensor
(hereafter EMT) for usual matter,
and $\:T^{ik}_{(de)}$ is the dark energy component,
which includes the famous cosmological constant.

From the Bianchi identity one gets the continuity equation
in the form:
 \begin{eqnarray}
 {T_{k}^{i}}_{~;~i}
= (T_{(m)~k}^{~~~~i} +  T_{(de)~k}^{~~~~~i})_{~;~i} = 0
 \label{divergence}
 \end{eqnarray}
where $T_{k}^{i}$ is the total EMT of the matter and dark energy.
In the case of non-interacting matter and dark energy the divergence
of each EMT equals zero separately. The general case of energy
transfer between matter and dark energy was considered by
Gromov et al. 2004 [11].

Note that gravity in GR is not at all equivalent to matter, 
so the total EMT  $\:T^{ik}$ does not contain the EMT of gravity field.
This is why Eq.(\ref{divergence}) is not a conservation
law for gravity plus total matter 
(Landau \& Lifshitz 1971 [19], sec.101, p.304).

\subsection{Einstein`s Cosmological Principle}

The second element of the SCM is the Einstein's Cosmological
Principle. This states that the universe is spatially homogeneous and
isotropic on "large scales" (see e.g.  Weinberg 1972 [31]; Peebles
1993 [23]; Peacock 1999 [22]).  Here the term "large scales" relates
to the fact that the universe is certainly inhomogeneous at scales of
galaxies and clusters of galaxies.  Therefore, the hypothesis of
homogeneity and isotropy of the matter distribution in space means
that starting from $r_{hom}$, at all scales $r>r_{hom}$ we can write
the total energy density $\varepsilon =\varrho c^2$ and the total
pressure $p$ as a function of time only:
\be
\label{rhohom}
\varepsilon(\vec{r},t) = \varepsilon(t)
\ee
\be
\label{preshom}
p(\vec{r},t)=p(t)
\ee
where the total energy density is the sum
of the energy density of ordinary matter 
($\varepsilon_{m}$) and dark energy ( $\varepsilon_{de}$),
and the total pressure is the sum of corresponding components:
\begin{eqnarray}
\label{eps2}
  \varepsilon =\varepsilon_{m} + \varepsilon_{de},\qquad
  p = p_{m} + p_{de}~~.
\end{eqnarray}
Here usual matter has equation of state
\begin{eqnarray}
 p_m = \beta\,\varepsilon_m,\qquad 0 \leq \beta \leq 1\,,
\label{pm}
\end{eqnarray}
and dark energy has equation of state
\begin{eqnarray}
 p_{de} = w\,\varepsilon_{de},\qquad
 -1 \leq w < 0\,.
\label{pq}
\end{eqnarray}
Recently values $w<-1$ also were considered.

With these equations of general relativity and expressions for
Cosmological Principle we are ready now to investigate the
properties of the standard cosmological model.

\subsection{Space expansion paradigm}

An important consequence of homogeneity and isotropy is that
the line element may be presented in the Robertson-Walker form:
\be
\label{rw1}
ds^{2} = c^{2}dt^{2} - S(t)^{2} d\chi^{2} - S(t)^{2} I_k(\chi)^{2}
(d\theta^{2}+\sin^{2}\theta d\phi^{2})]
\ee
where $\:\chi,\theta,\phi$ are the "spherical"
comoving space coordinates, $t$ is synchronous time coordinate,
$\:I_k(\chi) = \sin(\chi),\chi,\sinh(\chi)$
corresponding to curvature constant
values $\:k=+1,0,-1$ respectively and $\:S(t)$ is the scale factor.

In the {\em expanding space paradigm} the
proper metric distance $r$ of a  body with fixed comoving
coordinate  $\:\chi$ from the observer is given by
\be
\label{dist1}
r = S(t) \cdot \chi \ee and increases with time $\:t$ as the scale
factor $\:S(t)$.  Note that physical dimension of metric distance
$\:[r]=cm$, hence if $\:[S]=cm$ then $\:\chi$ is the dimensionless
comoving coordinate distance.  In fact $\chi$ is the spherical angle
and $S(t)$ is the radius of the sphere (or pseudosphere) embedded in
the 4-dimensional Euclidean space. 
It means that the "cm" (the measuring rod) itself is defined as 
unchangable unit of length in the embedding Euclidean space.
Hence the distance $r$ measured in $cm$ is the "internal" proper
distance on the 3-dimensional hypersurface of the embedding space.  In
other words $r$ and $\chi$ give the Eulerian and Lagrangian
representation of the comoving distance.

Often, "cylindrical" comoving space coordinates $\mu,\theta,\phi$ are
used in the literature. In this case the line element is
\be
\label{rw2}
 ds^{2} = c^{2}dt^{2} - S(t)^{2}
\frac{d\mu^{2}}{1-k\mu^2} - S(t)^{2} \mu^{2}
(d\theta^{2}+\sin^{2}\theta d\phi^{2})]
\ee
The metric distance $l$ is 
\be
\label{dist2}
l = S(t) \cdot \mu,
\ee
which can be interpreted as the "external"
distance from z-axis in an embedding Euclidean 4-dimensional space.
So it is important to use different designations for different
distances defined by intervals in Eq.\ref{rw1} and Eq.\ref{rw2}
(not as in Peacock 1999 [22], p.70).

The relation between these two
metric distances is
\be
\label{mdrel}
r = S(t) I^{-1}_k(l/S)
\ee
were $I^{-1}_k$ is the inverse function for $I_k$.

It is important to point out that the hypothesis of homogeneity of
space implies that for a given galaxy the recession velocity is
proportional to distance.  The {\em exact relativistic} expression for
the recession velocity $V_{exp}$ of a body with
fixed $\:\chi$, which due to the "space expansion" 
is the rate of increasing of the metric distance $r$ as a function of time, 
immediately follows from Eq.\ref{dist1} :
\be
\label{expvel}
V_{exp} = \frac{dr}{dt} =  \frac{dS}{dt} \chi = \frac{dS}{dt}
\frac{r}{S} = H(t) r = c \frac{r}{r_{H}}
\ee
where $\:H(t)= \dot{S}/S $
is the Hubble constant (also is a function of time)
and $\,r_{H} = c/H(t)$ is the Hubble distance at the time $\,t$.
(Here and in the following the dot indicates derivative 
with respect to the time $d/dt$.) 
This means that the
linear velocity-distance relation $V=Hr$, identified with the
observed Hubble law, can exist only if the matter distribution is uniform.  
However, according to modern data on galaxy distribution, this 
seems not to be the case at least for luminous matter. 

\subsection{Friedmann's equations}

In comoving coordinates the total EMT has the form:
\begin{equation}
T^k_i = diag \left(\varepsilon, -p, -p, -p \right)
\label{emt}
\end{equation}
In the case of unbounded homogeneous matter
distribution (Eqs.\ref{rhohom},\ref{preshom})
the Einstein's equations (Eq.\ref{eheq}) are directly
reduced to  the  Friedmann's equations. 
From the initial set of 16 equations we have only two  independent
equations for the (0,0) and (1,1) components,
which may be written in the following form:
\begin{eqnarray}
\frac{\dot S^2}{S^2} + \frac{kc^2}{S^2} = \frac{8\pi
G}{3c^2}~\varepsilon ~~,
\label{0.0}
\end{eqnarray}
\begin{eqnarray}
2\,\frac{\ddot S}{S} + \frac{\dot S^2}{S^2} + \frac{kc^2}{S^2} =
-\frac{8\pi G}{c^2}\,p ~~.
\label{1.1}
\end{eqnarray}
From the Bianchi identity (Eq.\ref{divergence}) it follows
the  continuity equation
 \begin{eqnarray}
 3 \frac{\dot S}{S} = -\frac{\dot \varepsilon}
 {\varepsilon + p}\,,
 \label{cont}
 \end{eqnarray}
which must be added to the Eqs.(\ref{0.0}), (\ref{1.1}) as
the consistency condition.

Using the definition of the Hubble constant $H=\dot{S}/S$,
we rewrite Eq.(\ref{0.0}) as
\begin{equation}
\label{friedmann00}
 H^2 - \frac{8\pi
 G}{3}\varrho=-\frac{kc^2}{S^2}\,,
\end{equation}
and equation (\ref{1.1}) as
\be
\label{friedmann11}
 \ddot{S}= - \frac{4 \pi G}{3} \left(
 \varrho+ \frac{3p}{c^{2}}\right)S\,.
\ee
In terms of the critical density $\varrho_{crit}=3H^2/8\pi G$, 
the total density parameter $\Omega = \varrho / \varrho_{crit}$, 
the curvature density parameter $\Omega_k = kc^2/S^2H^2$,
and the deceleration parameter $ q = -\ddot{S}S/\dot{S}^2 $, 
these equations also may be presented
in the form:
\begin{equation}
\label{f1}
 1 - \Omega = -\Omega_k\,,
\end{equation}
\be \label{f2}
 q = \frac{1}{2}\Omega \left(1 + \frac{3p}{\varrho c^2} \right)\,,
\ee
were  $\Omega, p ,\varrho$  are the total 
quantities, i.e. the sum of corresponding components for matter
and dark energy.

Solving the Friedmann's equation (Eq.\ref{f2})
one finds the dependence on time the scale factor $S(t)$
or the metric distance $\:r(t)$.

\section{Physics of space expansion}

\subsection{What does space expansion mean physically?}

The FLRW model gives an exact mathematical description of the expanding
space in the case of a geometrical  theory of gravity (GR). 
Increasing the scale factor $S(t)$ 
in FLRW metrics physically corresponds to expanding
space, that is adding vacuum, and homogeneous matter. 
Each comoving finite box in expanding universe 
continuously increases its volume,
so gets more and more cubic centimeters. 
Physically expansion of the universe  means 
the creation of space together with physical vacuum.
Creation of space may be visualized
by 2-d analogy with expanding sphere in 3-d space,
where the surface of the sphere increases with time and
for 2-d beings their universe grows with time
(gets more square centimeters) .

Real Universe is not homogeneous, it contains atoms, planets,
stars, galaxies. Bondi (1947) [5] considered spherical inhomogeneities
in the framework of GR
and showed that inside them the space expands slowly. In fact
bounded physical objects like particles, atoms, stars and galaxies
do not expands. So inside these objects there is no space creation.
This is why the creation of space is a new cosmological
phenomenon, which is not and cannot be tested in laboratory 
because the Earth, the Solar System and the Galaxy do not expand.

There are several
puzzling properties of FLRW models which are a direct consequence
of the above derived exact equations. 

\subsection{Violation of conservation laws in expanding space} 

Landau \& Lifshitz 1971 [19] in sec.101 "The energy-momentum pseudotensor"
emphasized that equation $T^i_{k~;~i}=0$ "does not generally express
a law of conservation", because of the mathematical structure
of the covariant divergence in Riemannian space. To get the total
(all kinds of matter plus gravity)
energy-momentum conserved, they suggest to consider energy-momentum 
pseudotensor, which could describe gravity itself.
However this violates the tensor character of the laws of conservation
and does not solve the problem 
of the energy density of the gravitational field
in a geometrical description of gravity. The root of the problem lies in
the equivalence principle and in the absence of a true gravity force in GR, 
while
all other fundamental fields have true forces, true EMTs and operate
in Minkowski space. It is important that Noether theorem relates
conserved EMT of material fields to maximal symmetry of the Minkowski space
and this is why in curved Riemannian space the EMT of gravity field
can not be properly defined.

The problem of the absence of true EMT for gravity field in cosmology
appears as the violation of energy conservation during the space
expansion. Indeed, let us consider the energy content of a comoving
ball with radius  $r(t)=S(t)\chi$. The volume element in metric
Eq.(\ref{rw1}) is
\begin{equation}
\label{dVprop}
  dV= S^3\,I_k^2(\chi)\,sin(\theta)d \chi
  d \theta \, d \phi ,
\end{equation}
and energy in the comoving sphere is
\begin{eqnarray}
  e(r) = \int_0^r T_0^0 dV =
  \frac{4\pi}{3}\,\varepsilon(t)\,S^3(t)\,\chi^3 \sigma_k(\chi),
\label{e1f}
\end{eqnarray}
where $\sigma_k(\chi) = \int\limits_0^\chi I_k^2(y)dy$,
so that it is equal to $1$ for
$k=0$, to $\,\, \frac{3}{\chi^3}\left(\frac{\chi^2}{2} - \frac{\sin
2\chi}{4} \right)$ for $k=1$, and to $\frac{3}{\chi^3}\left(\frac{\sinh
2\chi}{4}-\frac{\chi^2}{2} \right)$ for $k=-1$. 

To calculate the time dependence of the energy density we use 
the continuity equation (Eq.\ref{cont}) in the form
\begin{eqnarray}
 \dot{\varepsilon} = -3 (\varepsilon + p)\frac{\dot{S}}{S}\,~.
\label{contin}
\end{eqnarray}
For an ideal equation of state
 $p=\gamma \varrho c^2$ 
this equation has the simple solution 
\begin{equation}\label{rho1(s)}
  \varrho \propto  S^{-3(1+\gamma)}\,,
\end{equation}
where $S(t)$ is the scale factor.  
So in particular we have for dust,
radiation and vacuum
\begin{eqnarray}
\label{eps-2}
 \varrho_{dust} \propto  S^{-3}\,, \qquad
  \varrho_{rad} \propto  S^{-4} \,, \qquad
   \varrho_{vac}\propto const \,.
\end{eqnarray}
Hence the energy inside a comoving ball will change with time as
\begin{equation}\label{e-r-t}
  e(r) = \frac{4\pi }{3}\varrho c^2 r^3\sigma_k(\chi)
  \propto  S^{-3\gamma}(t)\,,
\end{equation}
so that for dust, radiation and vacuum we get
\begin{eqnarray}
\label{e-r}
 e_{dust}(r) \propto  const\,, \qquad
  e_{rad}(r) \propto  S^{-1} \,, \qquad
   e_{vac}(r)\propto S^{+3} \,.
\end{eqnarray}

Intriguingly the continuity equation (Eq.\ref{contin})
can be written also in the form
\begin{equation}\label{dE}
  dE + p~dV =0\,,
\end{equation}
where $dE = d(\varepsilon V) = d(\varrho c^2 V)$ 
is the change of energy within the
comoving volume $V = const \cdot S^3$. Interestingly, Eq.(\ref{dE})
looks like the law of conservation of energy in thermodynamics.
There is, however, an essential difference with the
cosmological case.

Eq.(\ref{dE}) in the laboratory means that if inside a finite box  
the energy decreases, it reappears outside the box as 
the work produced by the pressure acting on a piston of a
machine, increasing the volume of the box. The work performed
by the pressure inside the box is the cause of the 
energy decrease in the box.

In cosmology Eq.(\ref{dE}) gives us the possibility to calculate
of how much the energy increases or decreases inside a finite comoving
volume but it does not tell us where the energy comes from or where it goes. 
This is because the cosmological pressure does not produce work.
It was noted by Harrison(1981; 1995) [13, 15] that in a homogeneous
unbounded expanding FLRW model one may
imagine the whole universe partitioned into macroscopic cells,
each of comoving volume $V$, and all having contents in identical
states. The $-p\,dV$ energy lost from any one cell cannot reappear
in neighboring cells because all cells experience identical losses.
So the usual idea of an expanding cell performing work on its
surroundings cannot apply in this case. As Edward Harrison
emphasized:  "The conclusion, whether we like it or not, is obvious:
energy in the universe is not conserved" (Harrison, 1981 [13], p.276).

The same conclusion was reached by Peebles (1993) [23] when he considered
the energy loss inside a comoving ball of the photon gas (see our
Eq.\ref{e-r}). On page 139 he wrote "The resolution of this
apparent paradox is that while energy conservation is a good
local concept, ... there is not a general global energy conservation
in general relativity."

In fact, only for dust ($p=0$) one may speak about energy conservation
in expanding universe. But for any matter with $p\neq 0$ within any local
comoving volume energy is not conserved. This is because in GR there is
no EMT of gravity field and there is no gravity force in usual physical
sense.

\subsection{Newtonian form of the relativistic Friedmann equation }

Let us write Friedmann's Eq.(\ref{1.1}) in the form
\be
\label{freq-1}
\frac{d^{2}S}{dt^{2}}=
- \frac{4 \pi G}{3} S \left( \varrho+ \frac{3p}{c^{2}}\right)
\ee
Because of Lagrangian comoving coordinates do not depend on time,
one may rewrite Eq.\ref{freq-1} using Eq.\ref{dist1} as
\be
\label{freq2}
\frac{d^{2}r}{dt^{2}}= -\frac{GM_g(r)}{r^2}
\ee
where the gravitating mass $M_g(r)$ 
of a comoving ball with radius $r$
is given by
\be
\label{mg}
M_g(r) =
\frac{4\pi}{3} \left( \varrho + \frac{3p}{c^{2}}\right)r^3
\ee

Friedmann's equation (Eq.\ref{freq2}) in fact presents
the cosmological Friedmann force acting on a test galaxy with mass $m$
placed at distance $r$ from a fixed point at the center of
coordinate system:
\be
\label {frforce}
F_{Fr}(r) = m \frac{d^2r}{dt^2} = -\frac {GmM_g(r)}{r^2}
\ee
Therefore the exact relativistic equation describing the dynamical
evolution of the universe is exactly equivalent to the
non-relativistic Newtonian equation of motion of a test particle in
the gravity field of a finite sphere containing a mass 
$M_g$ within the radius
$r$. The second term in Eq.\ref{mg} does not change the Newtonian
character of the solutions.

Such a similarity was first mentioned by Milne(1934) [21] and
McCrea \& Milne(1934) [20], though they consider the Newtonian model an
approximation to Friedmann model. Later many authors claimed that the
Newtonian model can be used only for small radius compared to the
horizon distance. Here, however, we see that the Newtonian form of the 
Friedmann
equation is exact and true for all radius.  This creates a problem in
cosmology because Eq.\ref{frforce} places neither such relativistic
restrictions as motion velocity less than velocity of light, nor
retardation response effect.

The root of the puzzle lies in the geometrical description
of gravity in GR and in the
derivation of Friedmann's equation from Einstein's gravity
equations, using the
comoving synchronous coordinates with universal cosmic time $t$
and homogeneous unbounded matter distribution.

The Newtonian form of the Friedmann equation also creates the so
called {\em Friedmann-Holtsmark paradox.}  According to the Friedmann
equation there is the cosmological force Eq.(\ref{frforce}) acting on
a galaxy situated at the distance $r$ from another fixed galaxy.  This
is in apparent contradiction with well known Holtsmark result for the
probability density of the force acting between particles in infinite
Euclidean space in the case of $1/r^2$ behavior of the elementary
force (Holtsmark,1919 [16]; Chandrasekhar,1941 [6]).  For symmetry
reasons, due to the isotropy of the distribution of particles the
average force in any given location is equal to zero and one is left
with the finite value of fluctuating force, which is determined by the
nearest neighbor particles. Hence in infinite Euclidean space with
homogeneous Poisson distribution and Newtonian gravity force there is
no global expansion or contraction, but there is the density and
velocity fluctuations caused by local gravity force fluctuations.

Finally, the Newtonian form of the Friedmann equation explains why
recession velocities of distant galaxies can be larger
than the speed of light -- in Newtonian theory there is no
limiting velocity.
The exact relativistic velocity - distance  relation is the
Eq.\ref{expvel} and it is linear for all distances $\,r$. It means
that for $\,r>r_{H}$ we get $\,V_{exp}>c$ and the question arises
why general relativity violates special relativity. The usual answer
is that the space expansion velocity is not ordinary velocity of a
body in space, hence it has no ordinary limit by the velocity of
light. Again it demonstrates the unusual physics of the expanding
space.

\subsection{ Continuous creation of gravitating mass}

Puzzling properties of the FLRW model 
also come from consideration of the active
gravitating mass of the cosmological fluid, which may be 
either positive or negative and
changes sign with the cosmic time $t$. In the case of one fluid
with equation of state $p=\gamma \varrho c^2$ the active gravitating
mass (Eq.\ref{mg}) will be
\be \label{mgQ}
  M_{g}(r) = + \frac{4 \pi}{3}(1+3\gamma )\varrho r^3 \propto
  S^{-3\gamma }(t)\,.
\ee
So for dust, radiation and vacuum we get
\be
\label{mgdust}
  M_{dust}(r) =+ \frac{4\pi}{3}\varrho_{dust} r^3 \propto const(t) ,
\ee
\be \label{mgrad}
  M_{rad}(r)= +\frac{4\pi }{3}2\varrho_{rad} r^3 \propto S^{-1}(t),
\ee
\be \label{mgvac}
  M_{vac}(r) = - \frac{4 \pi}{3} 2\varrho_{vac} r^3 \propto
  -S^{+3}(t)\,.
\ee
Hence for the case of dust  the gravitating mass 
does not depend on time, 
but  in the case of radiation 
the gravitating mass continuously disappear in
the expanding universe. The most strange example is the vacuum, 
where the gravitating mass is negative (no such examples
in lab physics). 
This means that vacuum antigravity continuously increases in time
due to the continuous creation of gravitating (actually
"antigravitating") vacuum mass. In this sense the continuous
creation of matter in the Steady State cosmological model
is just a particular case of the new physics of the expanding
space.

\section{Cosmological redshift in expanding space}

Harrison (1981; 1993) [13, 14] clearly demonstrated that the cosmological
redshift due to the expansion of the 
universe is a new physical phenomenon and
is not the well known Doppler effect.
Recently this subject was intensively discussed by Kiang (2003) [18],
Davis \& Lineweaver (2003) [7] and Whiting (2004) [32] in an attempt to
clarify some "common big bang misconceptions" and the "expanding confusions"
widely spreaded in the literature.

\subsection{Lemaitre effect}

In the SCM the cosmological redshift is a new physical
phenomenon due to the expansion of space, which
induce the wave stretching of the traveling photons via the
Lemaitre's equation:

\be
\label{lem}
 (1+z) =
 \frac{\lambda_{0}}{\lambda_{1}} = \frac{S_{0}}{S_{1}}
\ee
where $z$ is cosmological redshift,
$\:\lambda_{1}$ and  $\:\lambda_{0}$ are the wavelengths at the
emission and reception, respectively,
and $\:S_{1}$ and $\:S_{0}$ the corresponding values of the scale
factor.  Equation (\ref{lem}) may be obtained from the radial
null-geodesics ($\:ds=0$, $\:d\theta=0$, $\:d\phi=0$) 
of the FLRW line element.

The cosmological redshift is caused by the Lemaitre effect, which
is different from the familiar Doppler effect. It is clear
from comparison between relativistic Doppler and cosmological FLRW
velocity-redshift V(z) relations.
To get V(z) in SCM one should consider first V(r) and r(z) relations.
Exact velocity-distance relation in FLRW model is Eq.\ref{expvel}:
\be
\label{v-r}
V_{exp} =  H(t) r(z) \,.
\ee
where $\:H(t)= \dot{S}/S $
is the Hubble constant at the time $\,t$.
The exact distance-redshift $r(z)$ relation in FLRW model is:
\begin{equation}\label{r-z}
  r(t_0,z)=r(z)= \frac{c}{H_0}\int_0^z\frac{dz'}{h(z')}\,,
\end{equation}
where $h(z)$ is taken from Friedmann equation (Eq.\ref{friedmann00})
\begin{equation}\label{h-z}
  h(z)=\sqrt{\tilde{\varrho}(z)\Omega^0 + (1-\Omega^0)(1+z)^2}\,,
\end{equation}
where $\Omega^0=\varrho^0_{tot}/\varrho^0_{crit}$ 
is the density parameter in present epoch,
$\tilde{\varrho}(z)=\varrho_{tot}/\varrho^0_{tot}$
is the normalized density of the total substances.

Analytical expressions for $r(z)$ may be obtained only in 
some simple cases. For the dust universe this relation was 
firstly derived by Mattig in 1958 and in terms of the internal
metric distance (Eqs.\ref{dist1}, \ref{mdrel}) it has the form:

\be \label{matt1}
   r(z)=S_0I_k^{-1}\left[\sqrt{\frac{2q_0 - 1}{k}}~
   \frac{zq_{0}+(q_{0}-1)
   ((2q_{0}z+1)^{1/2}-1)}{q_{0}^{2}(1+z)}\right]\,,
\ee

where scale factor $S(t=t_0)=S_0$ is

\be \label{s0}
  S_{0}=\frac{c}{H_{0}}\sqrt{\frac{k}{\Omega^0 - 1}}\,\,.
\ee

Now we can compare the exact FLRW $V_{exp}(z)$ relation with
exact relativistic Doppler $V_{Dop}(z)$ relation:
\begin{equation}\label{vexp(z)}
  V_{exp}(z) = c \frac{r(z)}{r_{Ho}}~~,
\end{equation}
\begin{equation}\label{vdop(z)}
  V_{Dop}(z)= c \frac{2z + z^2}{2 + 2z + z^2}~~.
\end{equation}
Clearly these are two different mathematical formulae which
corresponds to two different physical phenomena -- Lemaitre and
Doppler effects. Eqs.(\ref{vexp(z)},~\ref{vdop(z)})
give the same results only in the first order of $V/c$,
however the physics of space expansion is different from
motion in static space.

\subsection{Cosmological gravitational frequency shift}

In 1947 in the classic paper "Spherically symmetrical models in
general relativity" by Sir Hermann Bondi it was shown that,
at least for small redshifts,
the total cosmological redshift of a distant body may be expressed
as the sum of two effects:
the velocity shift (Doppler effect) due to the relative
motion of source and observer, and the global gravitational
shift (Einstein effect) due to the difference between the potential
energy per unit mass at the source and at the observer.
It means  that the spectral shift depend on the distribution of
matter in the space around the source. 
In the case of small distances
Bondi derived a simple formula for redshift which is simply
the sum of Doppler and gravitation effects, and which explicitly
showed that "the sign of the velocity shift depends on the sign
of $v$, but the Einstein shift is easily seen to be towards the red"
(Bondi,1947 [5],p.421).
Hence according to Bondi the cosmological gravitational frequency
shift is redshift (contrary to Peacock 1999 [22], p.619 and Zeldovich
\& Novikov 1984 [33] p.97 considerations). 

It was shown by Baryshev et al.(1994) [1] that from
Mattig's relation (Eq.\ref{matt1}) it follows directly for the case
of $z<<1, \,v/c \approx x = r/r_H$ that
\be
\label{zcos}
z_{cos}\approx x + \frac{1+q_0}{2}x^2 =
(\frac{v}{c} + \frac{1}{2}\frac{v^2}{c^2}) +
\frac{q_0}{2}x^2
\ee
is the sum of Doppler and gravitational redshifts:
\be
\label{zcos1}
z_{cos} \approx z_{Dop} + z_{grav}
\ee
where the cosmological gravitational redshift is
\be
\label{zgrav}
z_{grav} = \frac{\Delta\varphi(r)}{c^2} =
\frac{1}{2}\frac{GM(r)}{c^2 r} = \frac{1}{4}\Omega_0 x^2\,.
\ee
Note that the Eq.(\ref{zgrav}) describes the global gravitational
shift due to the whole mass within the ball having the center at the
source and the radius equal to the distance between the source and 
the observer. Hence cosmological gravitational shift
depends on the whole matter distribution between the source
and the observer (and should not be confused with the local 
gravitational shift at the source).

It is important that the center of the ball is placed at the source.
Then the cosmological gravitational redshift 
is consistent with the causality
principle according to which the event of emission
 of a photon by the source
(which marks the centre of the ball) must precede the event of
detection of the photon by an observer on the surface of the ball.
The detection event marks the
spherical edge of the ball, where all potential observers are situated.

In the literature there are a few discussions of the cosmological 
gravitational shift but they contain mistaken claimes. For instance,
if one consider the observer at the center of the cosmological ball
and a galaxy at the edge of the sphere, then one may conclude
that cosmological gravitational shift is blueshift (see Zeldovich \& \,
Novikov,1984 [33], p.97). Also one should use proper metric distance
for calculation the mass within a ball, instead of
angular distance used in Peacock,1999 [22], problem 3.4.

Note that in the case of the fractal matter distribution with fractal
dimension $D=2$ the cosmological gravitational redshift gives
the linear distance-redshift relation and becomes an observable
cosmological phenomenon (see e.g. Baryshev et al.1994 [1]).

\subsection{Hubble-deVaucouleurs paradox}

According to SCM the linear Hubble law is a consequence of the
homogeneity of the matter distribution.  However studies of the
3-dimensional local galaxy Universe have shown that at least in the
range of scales $\sim 1 \div 100 \;Mpc$ galaxy distribution is
strongly inhomogeneous and has fractal properties (Sylos Labini et
al.,1998 [25]; Baryshev \& Teerikorpi 2005 [4]).  This confirms de
Vaucouleurs' prescient view on the matter distribution so we call it
de Vaucouleurs law of large scale galaxy distribution (Baryshev et
al. 1998 [2]; Baryshev \& Teerikorpi 2002 [3]).

At the same time modern observations of the local Hubble flow based on
Cepheid distances to local galaxies, Tully-Fisher distances from the
KLUN program, and other distance indicators, demonstrate that the
linear Hubble law is well established within the
Local Volume ($r<10$ Mpc), starting from distances as small as
1 Mpc (see Teerikorpi,1997 [26]; Ekholm et al.,2001 [8];
Karachentsev et al. 2003 [17]; Teerikorpi et al. 2005 [27] ).

A puzzling conclusion is that the strictly linear
redshift-distance relation is observed
just inside inhomogeneous galaxy distribution, i.e. deep inside the
fractal structure for distances less than homogeneity
scale (it is known that $r_{hom} > 30$ Mpc):
\be
\label {hdev}
(\;r \;<\;r_{hom}\;)\;\;\& \;\;(\;cz=H_0r\;)
\ee

This empirical fact presents a profound
challenge to the standard model where the
homogeneity is the basic explanation of the Hubble law, and "the
connection between homogeneity and Hubble's law was the first success of
the expanding world model" (Peebles et al.,1991 [24]).
In fact, within the SCM one would not expect any neat relation of
proportionality between velocity and distance
for close galaxies, which are members of large scale structures.
However, contrary to the expectation, modern data show a good linear
Hubble law even for nearby galaxies. It leads to a new observationally
established puzzling fact that the linear Hubble law is not
a consequence of the homogeneity of visible matter, just because
the visible matter is distributed inhomogeneously.

The Hubble and de Vaucouleurs laws describe very different
aspects of the Universe, but both have in common universality and
observer independence.
This makes them fundamental cosmological laws and it is important
to investigate the consequences of their coexistence at
the same length-scales (see Baryshev et al.,1998 [2]; 
Gromov et al. 2001 [12]; 
Teerikorpi et al. 2005 [27]).

\section{Conclusion}

A cosmological model is in fact a particular solution of 
the gravity field equations. This is why
the roots of the conceptual problems of modern cosmology considered above 
actually lie in the theory of gravitation. In fact, 
all fundamental forces in physics (strong, weak, electromagnetic)
 have quantum nature,
(i.e. there are quanta of corresponding fields which carry
the energy-momentum of the physical interactions),
while GR is a non-quantum theory, which 
presents the geometrical interpretation of gravitational force
(i.e. the curvature of space itself 
which is not material field in space) 
and exclude the concept of localizable gravitational energy. 
This is why the main
problem of GR is the absence of the energy of the gravity field
or pseudo-tensor character of gravity EMT (Landau\& Lifshitz,1971 [19]).
Together
with GR the energy problem comes to cosmology and is the cause of
the conceptual problems of SCM.

It is also possible that in cosmology we see just one example of a new
physical phenomena where conservation laws are violated, receding
velocities of whole galaxies may exceed the velocity of light and
cosmological redshift is due to space expansion.  Note that the
explanation of the cooling of the photon gas in SCM, and hence the
origin of the cosmic microwave background radiation, rest on the
violation of the law of conservation of energy by the expanding
space. However physics of "space creation" is still not tested in
laboratory and hence needs more indirect observational evidence.

The big bang SCM is not the only possible model of the Universe.
There are several cosmological models which are based on other
fundamental hypotheses and give different interpretation of observable
phenomena.  A classification of possible relativistic cosmologies in
accordance with basic initial assumptions were discussed by Baryshev
et al.(1994) [1].  In particular relativistic quantum field approach
to gravity, where the Minkowski space and conservation laws are valid,
was considered by Feynman [9, 10]. Crucial observational tests of
alternative cosmological models and gravity theories should be
developed to understand real cosmological physics.

\begin{center}
{\bf Acknowledgments}
\end{center}

The author is grateful to the 
 Domingos Silva Teixeira
 for financial support.

\vspace{0.5cm}

\begin{center}
{\bf REFERENCES}
\end{center}


1. Baryshev, Yu., Sylos Labini F., Montuori, M., Pietronero, L.,
Facts and ideas in modern cosmology,
Vistas in Astronomy 38, 419, 1994.

2. Baryshev, Yu., Sylos Labini F., Montuori, M., Pietronero, L., Teerikorpi, P.,
On the fractal structure of galaxy distribution and its
implications for cosmology.
Fractals, vol. 6, 231, 1998. 

3. Baryshev Yu.V \& Teerikorpi P., \emph{ Discovery of cosmic fractals},
World Scientific, 2002.

4. Baryshev Yu.V \& Teerikorpi P., 
Fractal Approach to Large--Scale
Galaxy Distribution,
Bull. Spec. Astrophys. Obs. RAS, v.59, 2005 (astro-ph/0505185).

5. Bondi, H.,
Spherically symmetrical models in
general relativity,
 MNRAS 107, 410, 1947.

6. Chandrasekhar, S., Astrophys.J. 94, 511, 1941.

7. Davis T., Lineweaver C., 
Expanding confusion: common misconceptions of cosmological
horizons and the superluminal expansion of the universe, 2003
(astro-ph/0310808)

8. Ekholm, T., Baryshev, Yu., Teerikorpi, P., Hanski, M. \&
Paturel, G., 
On the quiescence of the Hubble flow in the vicinity of
the Local group:
a study using galaxies with distances from the Cepheid
PL-relation.
A \& A 368, L17, 2001.

9. Feynman R., \emph{Lectures on Gravitation} ,1962-63,
California Institute of Technology, 1971.

10. Feynman R., Morinigo F., Wagner W.,
 \emph{Feynman Lectures on Gravitation},
Addison-Wesley Publ. Comp., 1995.

11. Gromov A., Baryshev Yu., Teerikorpi P.,
 Two-fluid matter-quintessence FLRW models: energy transfer and
 the equation of state of the universe,
A \& A, 415, 813, 2004.

12. Gromov A., Baryshev Yu., Suson D., Teerikorpi P.,
Lemaitre-Tolman-Bondi model: fractality, non-simultaneous bang
time and the Hubble law,
Gravitation \& Cosmology, v.7, p.140, 2001.

13. Harrison E., \emph{ Cosmology: the science of the universe},
(Cambridge University Press), 1981
(2nd edition 2000)

14. Harrison, E. R.,
The redshift-distance and  velocity-distance laws,
ApJ 403,  28, 1993.

15. Harrison, E. R.,
Mining energy in an expanding universe,
ApJ 446, 63-66, 1995.

16. Holtsmark, J., Ann. d. Phys. 58, 577, 1919.

17. Karachentsev, I., Makarov, D.I., Sharina, M.E. et al., 
Local galaxy flows within 5 Mpc. 
A\&A 398, 479, 2003.

18. Kiang T.,  Time, distance, velocity, redshift:
a personal guided tour, 2003 (astro-ph/0308010)

19. Landau, L.D., Lifshitz, E.M.,\emph{ The Classical Theory of Fields},
(Pergamon Press), 1971.

20. McCrea W., Milne E.,  
Newtonian universes and the curvature of space,
Quart.J.Mathem., 5, 73, 1934.

21. Milne E., 
A Newtonian expanding Universe,
Quart.J.Mathem., 5, 64, 1934.

22. Peacock J.A.,  \emph{Cosmological Physics},
Cambridge Univ. Press, 1999.

23. Peebles P.J.E., \emph{ Principles of Physical Cosmology}
(Princeton Univ.Press), 1993.

24. Peebles P.J.E., Schramm D., Turner E., Kron R.,
The case for the relativistic hot Big Bang cosmology,
 Nature, 352, 769, 1991.

25. Sylos Labini F., Montuori, M., Pietronero, L.,
Scale-invariance of galaxy clustering,
Phys.Rep. 293, 61, 1998.

26. Teerikorpi P., Observational Selection Bias 
Affecting the
 Determination of the Extragalactic Distance Scale, ARAA 
35, 101, 1997.

27. Teerikorpi P., Chernin A., Baryshev Yu., 
The quiescent Hubble flow, local dark energy tests,
and pairwise velocity dispersion in a Omega=1 universe,
A \& A, 2005 (astro-ph/0506683).

28. Turner M., Making sense of the new cosmology, 2002a
(astro-ph/0202008).

29. Turner M.,  Dark Matter and Dark Energy: The 
Critical Questions,  2002b
(astro-ph/0207297).

30. Turner M., Absurd universe, Astronomy, November 2003, p.44

31. Weinberg S., \emph{ Gravitation and Cosmology}, (John Wiley \& Sons),
1972.

32. Whiting A., The expansion of space: free particle motion
and the cosmological redshift, 2004 (astro-ph/0404095).

33. Zeldovich Yu.B., Novikov I.D., \emph{ Relativistic Astrophysics}, Vol.2,
(The University of Chicago Press), 1984.


\end{document}